\documentclass[10pt,a4paper,twocolumn,english,pra,aps,showpacs,floatfix,groupedaddress,superscriptaddress]{revtex4-1}

\usepackage{graphicx}
\usepackage{epsfig}
\usepackage[english]{babel}
\usepackage{amsmath}
\usepackage{amssymb}
\usepackage{amsfonts}
\usepackage{longtable}
\usepackage{dcolumn}% Align table columns on decimal point
\usepackage{bm}
\usepackage{bbm}
\usepackage{color}
\usepackage{array}
\usepackage{colortbl}

\usepackage{bbold}
\usepackage{ctable}

\newcommand{\landau}[1]{\mathcal{O}\left(#1\right)}		%landau-O {term}
\newcommand{\I}{\mathbbm{1}}				% identity operator
\begin{document}
\title{Frequency modulated pulses for quantum bits coupled to time-dependent baths}

\author{Benedikt Fauseweh}
\email{benedikt.fauseweh@tu-dortmund.de}
\affiliation{Lehrstuhl f\"{u}r Theoretische Physik I, TU Dortmund, Otto-Hahn Stra\ss{}e 4, 44221 Dortmund, Germany}
\author{Stefano Pasini}
\email{s.pasini@fz-juelich.de}
\affiliation{Forschungszentrum J\"ulich, 52425 J\"ulich, Germany} 
\author{G\"otz S. Uhrig}
\email{goetz.uhrig@tu-dortmund.de}
\affiliation{Lehrstuhl f\"{u}r Theoretische Physik I, TU Dortmund, Otto-Hahn Stra\ss{}e 4, 44221 Dortmund, Germany}

\date{\rm\today}

\begin{abstract}
We consider the coherent control of a quantum bit by the use of short pulses with finite duration $\tau_\mathrm{p}$. 
By shaping the pulse, we perturbatively decouple the dynamics of the bath from the dynamics of the quantum bit during the pulse. Such shaped pulses provide single quantum bit gates robust against decoherence which are useful for quantum-information processing. We extend previous results in two ways: (i) we treat frequency modulated pulses
and (ii) we pass from time-independent baths to analytically time-dependent baths. First- and second-order solutions for $\pi$ and $\pi/2$ pulses are presented. They are useful in experiments where amplitude modulation is difficult
to realize.
\end{abstract}

\pacs{03.67.Pp, 82.56.Dj, 76.60.Lz, 03.65.Yz}

% 03.67.Pp Quantum error correction and other methods for protection against 
% decoherence (see also 03.65.Yz Decoherence; open systems; quantum statistical
%  methods; for decoherence in Bose-Einstein condensates, see 03.75.Gg) 
% 03.67.Lx Quantum computation
% 03.65.Yz Decoherence; open systems; quantum statistical methods 
% (see also 03.67.Pp in quantum information; for decoherence in Bose-Einstein 
% condensates, see 03.75.Gg) 
% 03.65.Vf Phases: geometric; dynamic or topological  
% 76.60.Lz Spin echoes
% 82.56.Dj High resolution NMR
% 82.56.Jn Pulse sequences in NMR
% 33.25.k Nuclear resonance and relaxation: Atomic and Molecular Physics

\maketitle

\section{Introduction}

The occurrence of decoherence in quantum systems is one of the main difficulties to be overcome in modern experiments. Especially quantum information processing (QIP) requires the quantum system to remain in fixed phase relations between the application of quantum gates. Otherwise one can not benefit from the quantum parallelism making QIP such a powerful technique.

A generic example for a quantum bit (qubit) is a spin $S=1/2$.  One of the first implementations of a quantum algorithm was realized with nuclear magnetic resonance (NMR) \cite{Vandersypen2001}. The qubits are encoded in the nuclear spin degrees of freedom. We also use the spin language to describe the states and dynamics of the qubit. The state $\uparrow$ is identified with the logical $|1\rangle$ and the state $\downarrow$ with the logical $|0\rangle$.

The loss of coherence is induced by the coupling of the spin to its environment, the so-called bath.
One way to suppress this coupling is the application of suitable control pulses first introduced by Hahn in 1950 
\cite{Hahn} for NMR experiments. This idea led to the development of sequences for control pulses ranging from the Carr-Purcell-Meiboom-Gill (CPMG) cycle \cite{carr54, meiboom:688} to more and more complex control schemes
\cite{haebe76}.
%\cite{PhysRevA.58.2733, PhysRevLett.94.060502, PhysRevLett.95.250501, PhysRevLett.95.180501, PhysRevLett.104.090501, PhysRevA.81.022309, luy05, skinn03, gersh07}. 
In QIP, this approach is known under the name of dynamic decoupling (DD) \cite{viola98,ban98,viola99a}. 
Theoretically, dynamic decoupling can be achieved to infinite order in the duration $T$ of the total pulse sequence
\cite{khodj05,khodj07}. One particularly efficient way to deal with pure dephasing decoherence is the use of theoretically optimized DD (Uhrig DD) \cite{uhrig07,uhrig08,yang08}. 
It has been successfully implemented experimentally \cite{bierc09a,du09}. 

The pulses considered in theoretical studies of DD schemes are mostly ideal in the sense that they have an infinite amplitude and that they act instantaneously in time corresponding to Dirac $\delta$ pulses. 
Of course, this property can not be realized in experiments. 
If the finite pulse duration is taken into account in calculations, it turns out to be a nuisance in most cases (see, for instance, Refs.\ \onlinecite{khodj07,pryad08b}). 
Hence, there is an abundant literature on pulse shaping and
optimization which we can mention only partly 
\cite{tycko83,levit86,cummi00,cummi03,skinn03,kobza04,skinn06,sengu05,motto06,alway07,pryad08a,pryad08b,pasin08a,pasin08b,pasin09a,khodj10} (for a book see Ref.\ \onlinecite{levit05}).
We emphasize, however, that a suitably shaped pulse can be integrated into
a DD sequence such that the high-order suppression of decoherence is hardly hampered \cite{uhrig10a,pasin11a}.

The majority of the existing theoretical studies of pulse shaping
consider pulses acting on the two-dimensional Hilbert space. The goal is to design robust pulses
which tolerate a maximum of frequency offset or other inaccuracies of the pulse \cite{tycko83,levit86,cummi00,cummi03,skinn03,kobza04,skinn06,alway07}. The next stage of complexity
includes random time-dependent classical noise, which is still described by classical fields coupled
to the spin \cite{motto06}. The maximum stage of complexity considers a fully quantum mechanical
bath which means that the qubit is coupled to a macroscopic quantum mechanical system by noncommuting
operators \cite{sengu05,pryad08a,pryad08b,pasin08a,pasin08b,pasin09a,khodj10}. It is on this level that
our present study is situated. We stress that a quantum mechanical pulse, which is robust
against a coupling to its environment, constitutes an appropriate single-qubit gate.

In particular, we extend previous work \cite{pasin09a} in two ways: (i) We allow for analytically 
time-dependent bath operators, both in the spin-bath coupling and in the Hamiltonian of the bath. 
Such time dependence may, for instance, arise from a time-dependent reference frame \cite{pasin10b}.
(ii) We propose frequency-modulated pulses while before only amplitude-modulated pulses
were studied \cite{sengu05,pryad08a,pryad08b,pasin08a,pasin09a,khodj10} except in the
general no-go theorem in Ref.\ \onlinecite{pasin08b}.
We stress that in the NMR context amplitude and phase-modulated pulses have been discussed
intensively \cite{skinn03,kobza04,skinn06}. But to our knowledge these investigations do not comprise
quantum mechanical baths nor dynamic classical noise.

Explicitly, we compute continuous solutions for $\pi$ and $\pi/2$ pulses realized by frequency modulation (see also Ref.\ \onlinecite{skinn06}). The consideration of frequency modulation is motivated from experimental situations where the frequency of a pulse can be controlled more accurately or more easily than its amplitude. 
Thus the present study is complementary to preceding ones.

The paper is organized as follows: In Sec.\ \ref{sec:model}, we give an overview of the model under study
and motivate our ansatz for the time evolution of the whole system. In Sec.\ \ref{sec:derivation} we derive the perturbative expansion for a generic time-dependent bath. We require that the time dependence is analytical in order to be able to apply a perturbative approach. Then we introduce the frequency-modulated ansatz in Sec.\ \ref{sec:freq} and specialize the general equations for this specific case. The solutions found for first- and second-order pulses are discussed in Sec.\ \ref{sec:results} and we finally conclude in Sec.\ \ref{sec:conclusion}.

\section{Model and Ansatz} 
\label{sec:model}

We consider the general case of a spin coupled to a time-dependent bath
\begin{align} \label{eq:model}
H(t) = H_{\mathrm{b}}(t) + \vec{\sigma} \cdot \vec{A}(t)  ,
\end{align}
where $H_\text{b}(t)$ denotes the part of the Hamiltonian that acts only on the bath. We refer to it as the
bath Hamiltonian. The vector of Pauli matrices $\vec{\sigma}$ acts on the Hilbert space of the spin $S=1/2$ 
while $\vec{A}(t)$ is a vector of bath operators to which the spin is coupled. No special operator structure is
assumed for the bath operators; i.e., the commutators $[ A_i(t'), A_i(t) ]$, $[ A_i(t'), A_j(t) ]$ and $[ A_i(t'), H_{\mathrm{b}}(t) ]$ do not need to vanish.

The model \eqref{eq:model} comprises typical cases such as a bosonic bath or a spin network.
Relevant experimental systems comprise the electronic spin in a quantum dot coupled to
the bath of nuclear isotope spins \cite{schli03} or the spin of a nitrogen vacancy center in diamond
interacting again with a  bath of nuclear isotope spins \cite{jelez06}.
For our purposes, we require that the time dependence of the operators $H_\text{b}(t)$ and $\vec{A}(t)$ 
is analytical so that they can be expanded in time:
\begin{subequations}
\begin{align}
\vec{A}(t) 		&= \vec{A}_0 		+ \vec{A}_1 t 		 + \vec{A}_2 t^2 	    + \cdots   , 
\label{eq:coef.A} \\
H_\mathrm{b}(t) &= H_{\mathrm{b},0} + H_{\mathrm{b},1} t + H_{\mathrm{b},2} t^2 + \cdots   . 
\label{eq:coef.H}
\end{align}
\end{subequations}
This analyticity is often fulfilled, e.g., in rotating reference frames or in the operator interaction picture. 
For fast time-dependencies, however, the above expansion is not useful because the
derivatives are large. Very fast oscillatory time-dependencies are better treated by
average Hamiltonian theory.

Our model includes the common case of a purely dephasing bath, i.e., a spin coupled only along the $\sigma_z$ direction to the bath. This model is justified in experiments where the dephasing time $T_2$ is 
significantly lower than the longitudinal relaxation time $T_1$. This is the case if the energetic splitting between the states with $\sigma_z=1$ and $\sigma_z = -1$ is large.

The coupling strength between the spin and the bath is given by $\lambda :=|| \vec{A}(t) || $
while the energy of the bath is defined to be $\omega_\mathrm{b} :=\left|\left| H_\mathrm{b}(t) \right|\right|$. If these operators are not bounded, that means if $\lambda$ and $\omega_\mathrm{b}$
cannot be defined by the operator norms, we refer by $\lambda$ and $\omega_\mathrm{b}$
to the generic energy scales of the corresponding operators. For instance, in a bosonic bath $\omega_\mathrm{b}$
is the upper cutoff of the bosonic energy spectrum.
The energy scales serve as reference values for $\tau_\mathrm{p}$. That means that
we aim at an expansion in the dimensionless ratios $\omega_\mathrm{b} \tau_\mathrm{p}$
and $\lambda\tau_\text{p}$.

Applying the control pulse to the system, the term 
\begin{align}
\label{eq:Hv}
H_v(t) = \vec{\sigma} \cdot \vec{v}(t)   
\end{align}
is added to the Hamiltonian $H(t)$. Here $\vec{v}(t)$ is a vector of amplitudes describing the controllable
time-dependent shape of the pulse. The normalized vector $\vec{v}(t)/\left| \vec{v}(t) \right|$
is the current axis of rotation at time $t$ while the norm $\left| \vec{v}(t) \right|$ describes the magnitude of
the control term which determines the velocity of rotation. 
Without loss of generality, we assume that the pulse starts at $t=0$ and ends at $t=\tau_\mathrm{p}$.
The time evolution between $0$ and $\tau_\mathrm{p}$ of the combined system reads
\begin{align}
U(\tau) &= T \left[ \exp \left( -i \int_0^{\tau} H(t) \mathrm{d} t - i \int_0^{\tau} H_v(t) \mathrm{d} t \right) \right]  
\end{align}
where $T$ stands for the standard time ordering.\\ Our aim is to perturbatively decouple the time evolution of the spin from the time evolution of the bath during the pulse. This motivates the following ansatz for the time evolution of the whole system
\begin{align}
 \label{eq:ansatz.tevolution}
U(\tau_\mathrm{p}) = U_\mathrm{b}(\tau_\mathrm{p}) P(\tau_\mathrm{p}) U_{\mathrm{c}}(\tau_\mathrm{p})   ,
\end{align}
where
\begin{subequations}
\begin{align}
U_\mathrm{b}(t) = T \exp \left( -i \int_0^{t} H_\mathrm{b}(t') \mathrm{d}t' \right) ,
\\
P(t) := T \exp \left( -i \vec{\sigma} \cdot \int_0^{t} \vec{v}(t') \mathrm{d} t'  \right) . 
\end{align}
\end{subequations}
The unitary operator $U_\mathrm{b}(\tau_\mathrm{p})$ describes the time evolution of the bath and
$P(\tau_\mathrm{p})$ the rotation of the spin due to the pulse. Note that the ansatz $U_\mathrm{b}(\tau_\mathrm{p}) P(\tau_\mathrm{p})$ does not comprise any coupling between spin and bath.
It is close to the goals of many previous studies aiming at robust pulses \cite{tycko83,levit86,cummi00,cummi03,skinn03,kobza04,skinn06,motto06,alway07}
and it corresponds to the ansatz used in previous studies separating
the pulse from a classcial \cite{motto06} or a quantum mechanical dynamics of the bath 
\cite{sengu05,pryad08a,pryad08b,pasin09a,khodj10}.
We emphasize that an ansatz which separates the pulse from the dynamics of the
spin plus bath system can be shown not to succeed beyond leading order \cite{pasin08a,pasin08b}.

Since the spin-bath coupling is not included in $U_\mathrm{b}(\tau_\mathrm{p}) P(\tau_\mathrm{p})$
we introduced the correction unitary operator $U_{\mathrm{c}}(\tau_\mathrm{p})$ in Eq.\ \eqref{eq:ansatz.tevolution}. We want to shape the pulse so that the correction term is as close to the identity as possible. A perfect decoupling would imply $U_{\mathrm{c}}(\tau_\mathrm{p})=\I$.
But this is unrealistic to achieve. Hence we pursue the perturbative approach to make as many terms of
an expansion in $\tau_\text{p}$ as possible vanish. Then
$U(\tau_\mathrm{p}) \approx U_\mathrm{b}(\tau_\mathrm{p}) P(\tau_\mathrm{p})$ represents
a valid approximation and one can neglect the spin-bath coupling during the pulse.
We remark that pulses shaped in this way constitute robust single-qubit gates.

\section{Derivation} 
\label{sec:derivation}

The derivation of the perturbative conditions for the shaped pulses is very similar to the derivation given in Ref.\ \onlinecite{pasin09a}. Yet we present a brief outline here in order to keep the present article self-contained
and because we extend the previous derivation to analytically time-dependent baths.
We start from the pulse Hamiltonian in Eq.\ \eqref{eq:Hv}. We describe the time-dependent pulse operator 
as a global rotation about the axis $\hat{a}(t)$
\begin{equation}
\label{eq:def.p}
P(t) = \exp \left( -i \vec{\sigma} \cdot \hat{a}(t) \frac{\psi(t)}{2} \right)  ,
\end{equation}
where $\left| \hat{a}(t) \right|= 1$. The spin is turned by the angle $\psi(t)$ at the time $t$.  Every unitary operator acting only on the Hilbert space of the spin can be written in the form of Eq.\ \eqref{eq:def.p}. In particular, a pulse that turns the spin by an angle $\chi$ satisfies
\begin{align} \label{eq:psi_taup}
 \psi(\tau_\mathrm{p}) = \chi . 
\end{align}
We stress the difference between the current axis of rotation $\vec{v}(t)/\left| \vec{v}(t) \right|$ and the effective axis $\hat{a}(t)$ describing the total rotation of the spin from its position at time $0$ to its  current position at time $t$.

By definition the pulse operator fulfills the Schr\"odinger equation
\begin{align}
i \partial_t P(t) = H_v(t) P(t).
\end{align}
which implies
\cite{pasin08b}
\begin{align}
2 \vec{v}(t) &=  \psi'(t) \hat{a}(t) +  \hat{a}'(t) \sin \psi(t) 
\nonumber \\
			 &- (1- \cos \psi(t))\left[ \hat{a}'(t) \times \hat{a}(t) \right] .
			 \label{eq:vdes} 
\end{align}
This differential equation is solved numerically for the frequency-modulated ansatz below.
The time evolution of the whole system is given by
\begin{align}
i \partial_t U(t) = \left[ H_\mathrm{b}(t) + \vec{\sigma} \cdot \vec{A}(t) + H_v(t) \right] U(t) .
\end{align}
Inserting the ansatz \eqref{eq:ansatz.tevolution} and solving for $\partial_t U_\mathrm{c}(t)$ yields
\begin{align}
\label{eq:uc.g2}
i \partial_t U_\mathrm{c}(t) &= G(t) U_\mathrm{c}(t) 
\\
\label{eq:uc.g}
G(t) &= P^{-1}(t) U_\mathrm{b}^{-1}(t) \, \vec{\sigma} \cdot \vec{A}(t) \, U_\mathrm{b}(t) P(t).
\end{align}
Thus the unitary correction is determined by a Schr\"odinger equation with $G(t)$ as its time-dependent Hamiltonian. The formal solution of Eq.\ \eqref{eq:uc.g2} in terms of  the standard time ordering operator
is
\begin{align}
\label{eq:uc.tordered}
U_\mathrm{c}(t) = T \left( \exp \left( -i \int_0^{t} G(\tau) \mathrm{d} \tau  \right) \right) .
\end{align}
Aiming at an expansion of $U_\mathrm{c}(t)$ in powers of $\tau_\text{p}$ it is convenient to 
 use the Magnus expansion \cite{MagnusExp} to express the time-ordered exponential
\begin{align}
\label{eq:magnus}
U_\mathrm{c}(\tau_\mathrm{p}) = \exp \left( - i \tau_\mathrm{p} \left( G^{(1)}+G^{(2)} + \dots \right) \right) ,
\end{align}
where $G^{(i)} = \landau{\tau_\text{p}^{i-1}}$. The first two terms read
\begin{subequations}
\label{eq:magnus_explicit}
\begin{align}
\tau_\mathrm{p} G^{(1)} &= \int_0^{\tau_\mathrm{p}} \mathrm{d}t G(t) \\
\tau_\mathrm{p} G^{(2)} &= -\frac{i}{2} \int_0^{\tau_\mathrm{p}} \mathrm{d} t_1 \int_0^{t_1} \mathrm{d} t_2 \left[G(t_1), G(t_2)\right]. 
\end{align}
\end{subequations}

Next, we need an expansion of $G(t)$ in powers of time. To this end, we consider
the representation \eqref{eq:def.p} which implies
\begin{subequations}
\begin{align} %Time dependences
P^{-1}(t) \vec{\sigma} \cdot \vec{A}(t) P(t) &= 
\left[ \cos \psi \vec{A} \right.  - \sin \psi \left( \hat{a} \times \vec{A} \right)   
			\nonumber \\
											 & + \left.(1-\cos \psi) \left(\hat{a} \cdot \vec{A} \right) \hat{a} \right] \cdot \vec{\sigma}   
			\label{eq:matrixcoeff}  \\
										  &= \vec{n}_{A(t)}(t) \cdot \vec{\sigma}    
			\label{eq:nA}\\
										  &= \sum_{i,j} n_{i,j}(t) A_j(t) \sigma_i ,
\end{align}
\end{subequations}
where the time dependencies on the right hand-side of Eq.\ \eqref{eq:matrixcoeff} are omitted to lighten the notation. The vector operator $\vec{n}_{A(t)}(t)$ is the vector $\vec{A}(t)$ after a rotation about the axis $\hat{a}(t)$ by the angle $-\psi(t)$. The corresponding rotation matrix $D_{\hat{a}}(-\psi)$ is given by its matrix elements $n_{i,j}(t)$; for their explicit form see Appendix \ref{app.c}. Due to the orthogonality of $D_{\hat{a}}(-\psi)$
the moduli of all its matrix elements are bounded by unity.

Note that there are two different kinds of time dependence in Eq.\ \eqref{eq:nA}.
On the one hand, the time dependence of $\vec{A}(t)$ becomes weaker and weaker as the pulse duration
$\tau_\text{p}$ is taken to zero because we assume that  $\vec{A}(t)$ is analytical. This is exploited below. On the other hand, the time dependence of the $n_{i,j}(t)$ scales with $\tau_\text{p}$,
which means that $\tilde n_{i,j}(s):=n_{i,j}(s\tau_\text{p})$ is completely independent of $\tau_\text{p}$
because the pulse is completed at $t=\tau_\text{p}$ whatever the pulse duration is.

We proceed by introducing the vector operator $\tilde{A}(t)$ and expanding it in powers of $t$
\begin{subequations}
\begin{align}
\tilde{A}(t) &:= U^{-1}_\mathrm{b}(t)A(t)U_\mathrm{b}(t)   
\\
&= \vec{A}_0 + it \left[ H_{\mathrm{b},0}, \vec{A}_0 \right] + t \vec{A}_1 + \landau{t^2}   .
\end{align}
\end{subequations}
Here the main differences to the derivation in Ref.\ \onlinecite{pasin09a} arises.
In Ref.\ \onlinecite{pasin09a}, the term proportional to $\vec{A}_1$ did not appear because the bath was considered to be time-independent. Using the vector operator $\vec{n}_{A(t)}(t)$ from Eq.\ \eqref{eq:nA} we rewrite $G(t)$ concisely as
\begin{subequations}
\begin{align}
G(t) &= P^{-1}(t) \vec{\sigma} \cdot \tilde{A}(t) P(t)   
\\
	 &= \vec{n}_{\tilde{A}(t)}(t) \cdot \vec{\sigma}  .
	 \label{eq:gbynA}
\end{align}
\end{subequations}
This form of $G(t)$ can be expanded in powers of $t$ such that the
neglected terms are of second order in $\tau_\text{p}$ for $t\in[0,\tau_\text{p}]$
\begin{align}
\label{eq:Gexpand}
G(t) = \vec{n}_{A_0} \cdot \vec{\sigma} + t \left(i \left[ H_{\mathrm{b}, 0}, i \vec{\sigma} \cdot \vec{n}_{A_0} \right] +  \vec{n}_{A_1} \cdot \vec{\sigma} \right) + \landau{\tau_\text{p}^2}.
\end{align}
Note that the time dependence stemming from the pulse rotation is not expanded 
because it does not change on $\tau_\text{p}\to 0$. Physically this means that
one can expand in $\tau_\mathrm{p} H(t)$, i.e., in $\lambda\tau_\mathrm{p}$
and in $\omega_\text{b}\tau_\mathrm{p}$, but \emph{not} in $\tau_\mathrm{p} H_v(t)$
because the magnitude of $H_v(t)$ has to be increased on $\tau_\text{p}\to 0$
to realize the desired pulse.

Inserting Eq.\ \eqref{eq:Gexpand} in the terms of the Magnus expansion \eqref{eq:magnus_explicit}
eventually yields
\begin{subequations}
\begin{align}
U_\mathrm{c}(\tau_\mathrm{p}) &= \exp \left( -i \left( \eta^{(1)} + \eta^{(2)} + \dots \right) \right)  ,
\\
\eta^{(1)} &= \underset{i,j}{\sum} \sigma_i A_{j,0} \int_0^{\tau_\mathrm{p}} n_{i,j}(t) \mathrm{d} t  \label{eq:eta1} ,  \\
\eta^{(2)} &= \underset{i}{\sum} \sigma_i \left( \eta_i^{(2a)}+\eta_i^{(2b)} \right) + \eta^{(2c)}  \label{eq:eta2}  ,
\end{align}
\end{subequations}
where $\eta^{(i)}\propto \tau_\text{p}^i$. Explicitly, one has
\begin{subequations}
\label{eq:eta2_detail}
\begin{align}
\eta_i^{(2a)} &= \underset{j}{\sum} \left( \left[H_{\mathrm{b},0},A_{j,0}\right] -i A_{j,1}  \right) \int_0^{\tau_\mathrm{p}} t \, n_{i,j}(t) \mathrm{d} t ,
\label{eq:eta2a}    \\
\eta_i^{(2b)} &= \underset{l,m}{\sum} \left[A_{l,0},A_{m,0}\right]_{+} \int_0^{\tau_\mathrm{p}} \mathrm{d} t_1 \int_0^{t_1} \mathrm{d} t_2  \nonumber  \\ 
					   &\cdot \underset{j,k}{\sum} \epsilon_{ijk} n_{j,l}(t_1) n_{k,m}(t_2)    ,
\label{eq:eta2b}\\
\eta^{(2c)} &= \underset{i,j<k}{\sum} \left[A_{j,0},A_{k,0}\right] \int_0^{\tau_\mathrm{p}} \mathrm{d} t_1 \int_0^{t_1} \mathrm{d} t_2 \nonumber \\
		    &\cdot \left( n_{i,j}(t_1) n_{i,k}(t_2) - n_{i,j}(t_2) n_{i,k}(t_1) \right)   
		    \label{eq:eta2c}.
\end{align}
\end{subequations}
In these equations the anticommutator $\left[ \cdot , \cdot \right]_+$ and the completely antisymmetric Levi-Civita tensor $\epsilon_{ijk}$ appear. The indices $i,j,k,l,$ and $m$ take one of the values $x,y,$ or $z$.
Most of the Eqs. \eqref{eq:eta1} and \eqref{eq:eta2_detail} are identical to those obtained in Ref.\ \onlinecite{pasin09a}. Only in $\eta_i^{(2a)}$ does the time dependence of the bath appear additionally.
It is encoded in the operators $A_{j,1}$ which are zero for a time-independent bath.

The second-order equations \eqref{eq:eta2_detail} do not show a dependence on pure bath terms of an order higher than $H_{\mathrm{b},0}$. This means that the time dependence of the pure bath Hamiltonian is irrelevant up to second order. This is reasonable because even for a time-independent bath the actual bath dynamics induced from $H_\mathrm{b}$  appears only in second-order conditions. This does not apply to the vector operator $\vec{A}(t)$ which is already relevant in first-order pulses.

In the general case of a completely generic bath, all the expressions $\eta^{(\alpha)}$, 
$\alpha\in\{1,2a,2b,2c\}$, have
to vanish in order to fulfill $U_\mathrm{c}=\I +\landau{\tau_\text{p}^3}$. The pulse shape determines the time
 evolution of the matrix elements $n_{i,j}$. Hence, in order to fulfill all the conditions the operator-independent integrals in Eqs.\ \eqref{eq:eta1} and \eqref{eq:eta2_detail} must disappear. The resulting 39 scalar equations are identical to  those obtained in Ref.\ \onlinecite{pasin09a}.

This is our first key result. It proves the applicability of the previously obtained pulses even in the
presence of a non-trivial time dependence of the bath which may stem from special reference frames or from the interaction picture of fast modes. For specific cases, such as the pure dephasing model or if $[H_\mathrm{b}(t), \vec{A}(t)] = 0$ the number of scalar equations to be fulfilled for $U_\mathrm{c}=\I +\landau{\tau_\text{p}^3}$ is reduced significantly. Pulses with less complexity can be used. This is studied in the sequel.

\section{Frequency modulated ansatz} 
\label{sec:freq}

\begin{table}
\begin{ruledtabular}
 \begin{tabular}{lrlrlr}
\multicolumn{4}{c}{First-order pulses}  \\
%\hline
\multicolumn{2}{c}{FM-1-PI} &\multicolumn{2}{c}{FM-1-PI2} \\
\hline                         
 $V_0$ &  3.75146609 & $V_0$ &  4.92892484 \\
 $b_1$ &  0.00011442 & $b_1$ &  0.00009874 \\
 $b_2$ & -1.09347112 & $b_2$ &	-0.94331659	\\
 $b_3$ &  0.00012443 & $b_3$ &	 0.00002530 \\
 $b_4$ & -0.59452572 & $b_4$ &	-0.12087663	\\                     
\end{tabular}
\end{ruledtabular}
\caption{Overview of the pulses satisfying all first-order equations \eqref{eq:eta1}. FM-1-PI denotes the frequency modulated $\pi$ pulse. FM-1-PI2 denotes the frequency-modulated $\pi/2$ pulse. The dimensionless coefficients $b_n$ belong to the ansatz in Eq.\ \eqref{eq:omega_asy}. The amplitudes $V_0$ are given in units of $1/\tau_\mathrm{p}$. 
With all eight digits given the conditions are fulfilled for the $\pi$ pulse within $10^{-10}$
and for the $\pi/2$ pulse within $10^{-9}$. With only two digits
they are fulfilled within $10^{-3}$ and $10^{-2}$, respectively.
\label{tab.first_order}}
\end{table}

\begin{table}
\begin{ruledtabular}
 \begin{tabular}{lrlr}
\multicolumn{4}{c}{Second-order pulses}  \\
%\hline
\multicolumn{2}{c}{FM-2-PI} &\multicolumn{2}{c}{FM-2-PI2} \\
\hline								    
$V_0$ & 	12.83432979 & $V_0$ 		& 12.25619390	\\
$b_1$ &  	 0.11475139 & $b_1$ 		&  1.73071840 	\\
$b_2$ &  	 0.17248587 & $b_2$ 		&  0.73529959 \\
$b_3$ &  	 0.48262521 & $b_3$ 		&  0.23242523 \\
$b_4$ &     -1.14494851 & $b_4$ 		& -0.24829310 \\
$b_5$ & 	-0.20879091 & $b_5$ 		& -0.07102204 \\
$b_6$ &  	 0.25378013 & $b_6$ 		& -0.13192380 \\
$b_7$ &  	 0.20306835 & $b_7$ 		&  1.07948226 \\
$b_8$ & 	-0.16748022 & $b_8$ 		&  0.12220006 \\
$b_9$ & 	-0.32052254 & $b_9$ 		&  0.04608986 \\
$b_{10}$ & 	 0.32586203 & $b_{10}$ 		& -0.15365617 
\end{tabular}
\end{ruledtabular}
\caption{Overview of the pulses satisfying all first- and second-order equations, Eqs.\ (\ref{eq:eta1},\ref{eq:eta2_detail}). FM-2-PI and FM-2-PI2 denote the frequency-modulated $\pi$ pulse and $\pi/2$ pulse, respectively. The dimensionless coefficients $b_n$ belong to the ansatz in Eq.\ \eqref{eq:omega_asy}. The amplitudes $V_0$ are given in units of $1/\tau_\mathrm{p}$. 
With all eight digits given the conditions are fulfilled for the $\pi$ pulse within $10^{-10}$
and for the $\pi/2$ pulse within $10^{-11}$. With only two digits
they are fulfilled within $10^{-2}$ and $10^{-2}$, respectively.
\label{tab.second_order} }
\end{table}

\begin{table}
\begin{ruledtabular}
 \begin{tabular}{lrlr}
\multicolumn{4}{c}{Minimized second-order pulses}  \\
%\hline 
\multicolumn{2}{c}{FM-2-MIN-PI} &\multicolumn{2}{c}{FM-2-MIN-PI2} \\ 
\hline	                
$V_0$ &    10.70711454		& $V_0$    &  8.43541412		\\
$b_1$ & 	0.00002087 	 	& $b_1$    & -1.82041507 		\\
$b_2$ & 	1.38768938 	 	& $b_2$    & -0.35249197	 	\\
$b_3$ &    -0.00019922 	 	& $b_3$    &  0.03054874	 	\\
$b_4$ &    -0.70668998    	& $b_4$    &  0.52093576	 	\\
$b_5$ &    -0.00001588		& $b_5$    & -0.55504440	 	\\
$b_6$ & 	0.13773085 	 	& $b_6$    & -0.38815568	 	\\
$b_7$ & 	0.00008770 		& $b_7$    &  0.45167361	 	\\
$b_8$ & 	0.68894331		& $b_8$    & -0.19445080	 	\\
$b_9$ &    -0.00011408		& $b_9$    & -0.16194806   		\\
$b_{10}$ & -0.69744086		& $b_{10}$ & -0.28223330	 	\\
$b_{14}$ &  0.46501991	 	& $b_{14}$ &  0.04585897	 
\end{tabular}
\end{ruledtabular}
\caption{
Overview of the pulses satisfying all first- and second-order equations, Eqs.\ (\ref{eq:eta1},\ref{eq:eta2_detail}) with minimized
amplitude. FM-2-PI and FM-2-PI2 denote the frequency-modulated $\pi$ pulse and $\pi/2$ pulse with minimized
amplitude, respectively. The dimensionless coefficients $b_n$ belong to the ansatz in Eq.\ \eqref{eq:omega_asy}. The amplitudes $V_0$ are given in units of $1/\tau_\mathrm{p}$. 
With all eight digits given the conditions are fulfilled for the $\pi$ pulse within $10^{-10}$
and for the $\pi/2$ pulse within $10^{-9}$. With only two digits
they are fulfilled within $10^{-3}$ and $10^{-1}$, respectively.
\label{tab.second_order_min} }
\end{table}

\begin{figure}
 \includegraphics[width=\columnwidth]{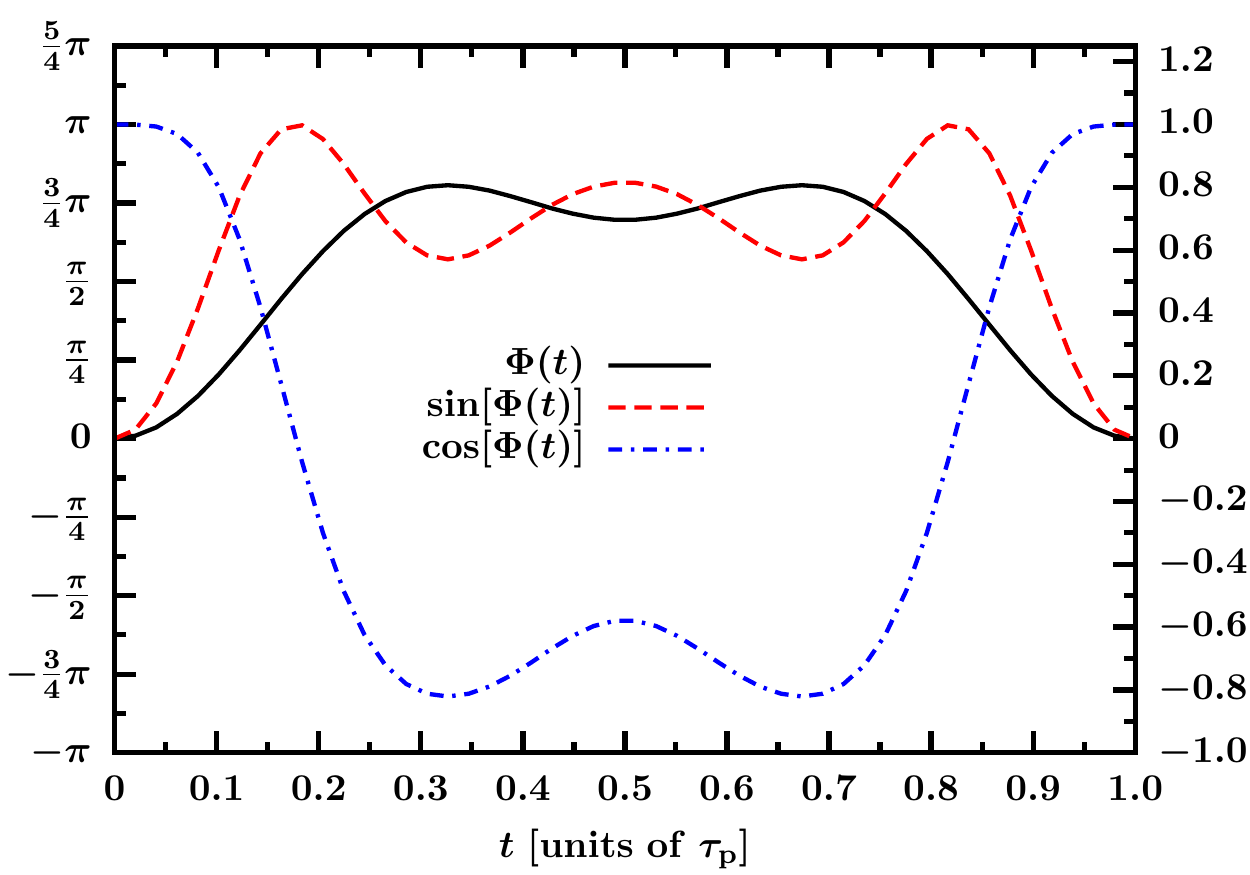}
 \caption{(Color online) First-order $\pi$ pulse FM-1-PI. We also plot $\sin \Phi(t) \propto v_y(t)$ and $\cos \Phi(t) \propto v_x(t)$ to illustrate the pulse shape in spin space. The left scale refers to $\Phi(t)$ and the right scale to $\sin \Phi(t)$ and $\cos \Phi(t)$, respectively. The Fourier coefficients for this pulse are given in Table \ref{tab.first_order}. 
 \label{fig.FC-1ASY-PI}}
\end{figure}

To solve Eq.\ \eqref{eq:vdes} we choose an ansatz for $\vec{v}(t)$. In this paper we focus on a frequency-modulated pulse acting only in the $\sigma_x \sigma_y$-plane with a fixed amplitude $V_0:=\left|\vec{v}\right|$ and the current axis of rotation
\begin{align}
\label{eq:ansatz}
\vec{v}(t) = \left(\begin{array}{c}V_0 \cdot \cos(\Phi(t))
\\
V_0 \cdot \sin(\Phi(t))\\0\end{array}\right),
\end{align}
where $\Phi(t)$ is a time-dependent phase. Note the difference to pulses with a time-dependent amplitude and a fixed axis discussed, for instance, in Refs.\ \onlinecite{pryad08a,pryad08b,pasin09a}. 
We focus here on frequency modulation in complement to previous work because there may be experimental setups where frequency modulation is much easier (or more accurately) implemented than amplitude modulation.
Note that the ansatz \eqref{eq:ansatz} assumes that the control pulse can be
switched on instantaneously. Transients are assumed to be sufficiently steep to be taken as jumps.
The consideration of continuous amplitudes and frequency modulation is left to future research.

To point out the relation of the ansatz \eqref{eq:ansatz} to the experimental realization in
the laboratory framework we consider a spin with a Larmor frequency $\omega_\mathrm{L}$ in the NMR language  \cite{RevModPhys.76.1037}:
\begin{align}
H_z = - \frac{\omega_\mathrm{L}}{2} \sigma_z .
\end{align}
Of course, this description is not restricted to nuclear spins. 
Any two-level system with an energy splitting can be considered. 
The control field is realized by applying a field perpendicular to the $\sigma_z$-axis 
rotating with the Larmor frequency:
\begin{align}  
\label{eq:fm}
H_\mathrm{rf} = V_0 \left\lbrace \sigma_x \cos \left[\omega_\mathrm{L} t - \Phi(t)\right]
 - \sigma_y \sin \left[\omega_\mathrm{L} t - \Phi(t)\right]\right\rbrace.
\end{align}
We include a time-dependent phase $\Phi(t)$ to shape the pulse. Its derivative $\partial_t \Phi(t)$ is the deviation of the frequency from the Larmor frequency.
In this sense Eq.\ \eqref{eq:fm} describes a frequency-modulated pulse.
Next, $H_\mathrm{rf}$ is transformed into the rotating framework in which
$H_z$ vanishes. Using the unitary time evolution induced by $H_z$
\begin{align}
U_\mathrm{rot}(t) = \exp \left( i \frac{\omega_\mathrm{L}}{2} t \sigma_\mathrm{z} \right) ,
\end{align}
we obtain $H_\mathrm{rot}(t)=U^\dag_\mathrm{rot}H_\mathrm{rf}U_\mathrm{rot}$,
which reads
\begin{align}
H_\mathrm{rot}(t) = \left(\begin{array}{c}V_0 \cdot \cos(\Phi(t))\\V_0 \cdot \sin(\Phi(t))\\0\end{array}\right) \cdot 		  	 
				 \left(\begin{array}{c}\sigma_x		\\ \sigma_y		\\ \sigma_z\end{array}\right) = \vec{v}(t) \cdot \vec{\sigma} .
\end{align}

In order to find $\hat{a}(t)$ and $\psi(t)$ appearing in the parametrization in Eq.\ \eqref{eq:def.p}  of the pulse
one has to solve the differential equation \eqref{eq:vdes}. 
Because $\hat{a}(t)$ is a unit vector, it is convenient to describe it by two angles:
$\varphi(t)$ and $\theta(t)$
\begin{align}
\label{eq:a_para}
\hat{a}(t) = \left(\begin{array}{c}\sin(\theta(t)) \cos(\varphi(t))\\\sin(\theta(t)) \sin(\varphi(t))\\\cos(\theta(t))\end{array}\right).
\end{align}
Solving Eq.\ \eqref{eq:vdes} for the time derivatives of $\psi(t)$, $\varphi(t)$, and $\theta(t)$,
we find
\begin{subequations}
\label{eq:des_result}
\begin{align}
\partial_t \psi &= 2 V_0 \sin \theta \left[ \sin \Phi \sin \varphi + \cos \Phi \cos \varphi \right] ,    \label{eq:des_result1}\\
\partial_t \varphi &= V_0 \frac{\left[ \cos \frac{\psi}{2} \sin (\Phi - \varphi) - \sin \frac{\psi}{2} \cos \theta \cos (\Phi - \varphi) \right]}{\sin \frac{\psi}{2} \sin \theta} ,
\label{eq:des_result2} \\
\partial_t \theta &= V_0 \frac{\left[ \cos \frac{\psi}{2} \cos \theta \cos (\Phi - \varphi) + \sin \frac{\psi}{2} \sin (\Phi - \varphi) \right]}{\sin \frac{\psi}{2}}.
  \label{eq:des_result3}
\end{align}
\end{subequations}
The seeming singularities for vanishing angles on the right-hand sides of Eqs.\ \eqref{eq:des_result2} and \eqref{eq:des_result3} have no \emph{physical} reason, but they 
only result from the choice of spherical coordinates and from the chosen parametrization in Eq.\ \eqref{eq:def.p}.
Note that the global axis of rotation $\hat a$ is not uniquely defined if
$\psi$ is a multiple of $2\pi$.

At the very beginning at $t=0$ the current axis of rotation $\vec{v}$ and the global one $\vec{a}$
coincide. The former lies by construction in the $\sigma_x \sigma_y$-plane. Hence we have the initial conditions
\begin{subequations}
\begin{align}
\underset{t \to 0}{\lim} \theta(t) &= \frac{\pi}{2},\\
\underset{t \to 0}{\lim} \psi(t) &= 0,\\
\underset{t \to 0}{\lim} \varphi(t) &= \Phi(0) ,
\end{align}
\end{subequations}
where the latter two equations represent our deliberate choice.
Inspecting the limit $t\to 0$ one additionally finds
\begin{subequations}
\begin{align}
2 \partial_t \varphi \mid_{t = 0} &= \left. \partial_t \Phi(t) \right|_{t = 0}  , \\
\partial_t \theta \mid_{t = 0} &= 0  .  
\end{align}
\end{subequations}
The derivative $\partial_t\psi$ follows trivially from Eq.\ \eqref{eq:des_result1}.
In the next section, we provide solutions for this ansatz and a specific case of spin-bath coupling.

\section{Results} 
\label{sec:results}

\begin{figure}
 \includegraphics[width=\columnwidth]{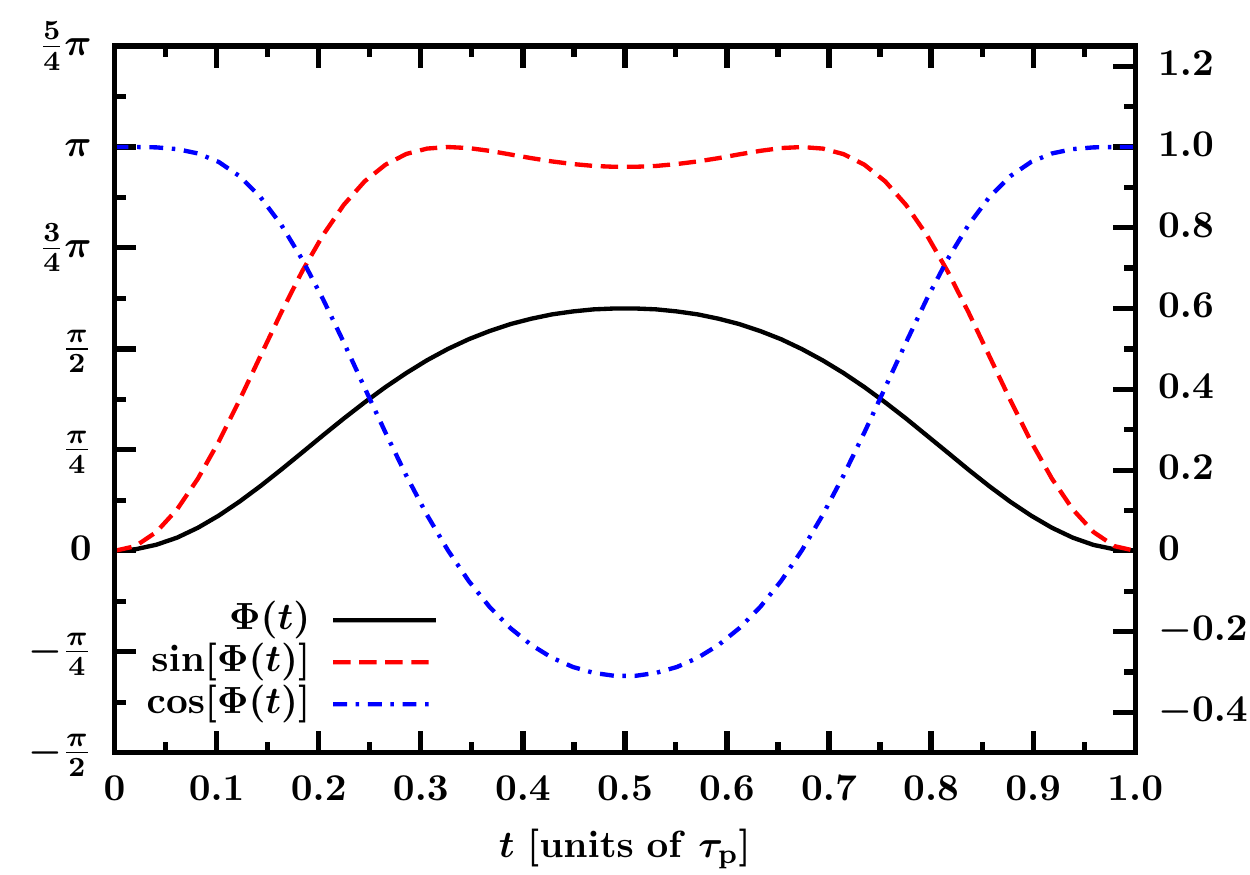}
  \caption{(Color online) First-order $\pi/2$ pulse FM-1-PI2. We also plot $\sin \Phi(t) \propto v_y(t)$ and $\cos \Phi(t) \propto v_x(t)$ to illustrate the pulse shape in spin space. The left scale refers to $\Phi(t)$ and the right scale to $\sin \Phi(t)$ and $\cos \Phi(t)$, respectively. The Fourier coefficients for this pulse are given in Table \ref{tab.first_order}. 
  \label{fig.FC-1ASY-PI2}}
\end{figure}

We are interested in the experimentally important case of a purely dephasing model, i.e., a bath coupled only via $\sigma_z$ to the spin
\begin{align}
\label{eq:pure_dephas}
\vec{A}(t) = A(t)\vec{e}_z.
\end{align}
Hence the coupling becomes simpler, but the bath dynamics itself is still kept in full generality. 
Spin flips do not occur in this model so that $T_1$ is infinite. But decoherence of the $T_2$ type is entirely kept.  This assumption is justified in many experimental realizations. Moreover, the
simplification of the coupling is advantageous for pulse shaping because it
reduces the number of integral conditions derived from Eqs.\ \eqref{eq:eta1} and
 \eqref{eq:eta2_detail} in second-order to be fulfilled from 39 to
3 first-order conditions and 6 second-order conditions which are given explicitly in Appendices
 \ref{app.a} and \ref{app.b}.

\begin{figure}
 \includegraphics[width=\columnwidth]{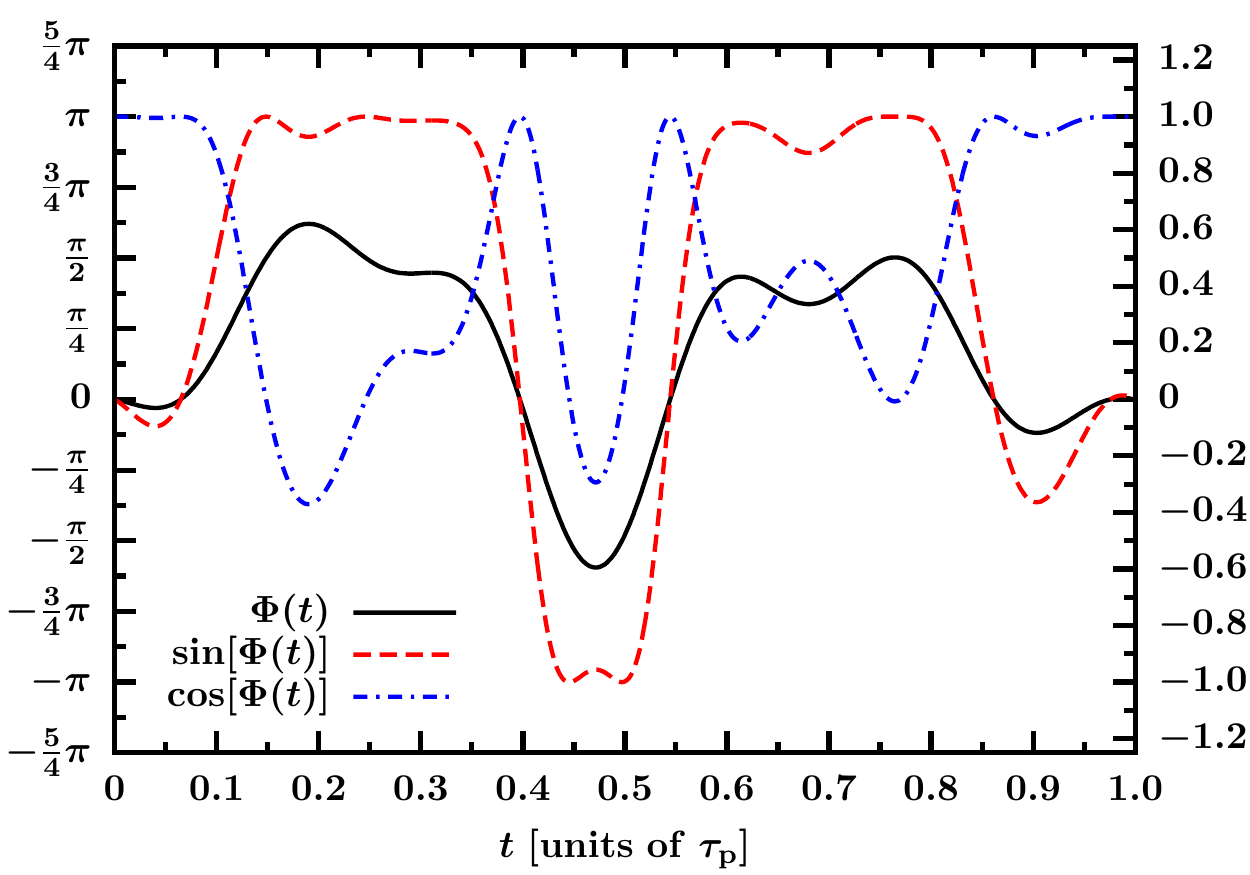}
 \caption{(Color online) Second-order $\pi$ pulse FM-2-PI. We also plot $\sin \Phi(t) \propto v_y(t)$ and $\cos \Phi(t) \propto v_x(t)$ to illustrate the pulse shape in spin space. The left scale refers to $\Phi(t)$ and the right scale to $\sin \Phi(t)$ and $\cos \Phi(t)$, respectively. The Fourier coefficients for this pulse are given in Table \ref{tab.second_order}.
 \label{fig.FC-2ASY-PI}}
\end{figure}

In the following, we present continuous pulses which fulfill the first-order conditions (first-order pulses) and
 pulses which fulfill all first- and second-order conditions (second-order pulses) for pure dephasing as in Eq.\ \eqref{eq:pure_dephas}. Thereby, we provide optimized pulses that decouple the spin from the bath during the duration of the pulse up to  $\landau{\tau_\mathrm{p}^3}$.

In order to consider a continuous frequency modulation we 
use the Fourier series ansatz 
\begin{align}
\Phi(t) &= {\underset{n}{\sum}} b_{2n-1} \sin \left( 2 \pi n {t/\tau_p} \right) 
					   + 					   b_{2n} \left[\cos \left( 2 \pi n {t/\tau_p} \right)- 1 \right] 
					   \label{eq:omega_asy} 
\end{align}
for $\Phi(t)$.
We consider $\pi$ and $\pi/2$ pulses because of their frequent use in QIP and NMR. Therefore, the pulse has to fulfill 
\begin{equation}
\label{eq:angle}
\psi(\tau_\mathrm{p}) = \pi \quad \text{or} \quad \pi/2
\end{equation}
for $\pi$ pulses and $\pi/2$ pulses, respectively, according to Eq.\ \eqref{eq:psi_taup}. 
The value $\theta(\tau_\mathrm{p})$ is fixed by the fact that the final axis of rotation has to be 
perpendicular to $\sigma_z$ to rotate the spin by the full angle $\psi(\tau_\mathrm{p})$. Thus we require
\begin{align} 
\label{eq:theta_taup}
\theta(\tau_\mathrm{p}) = \frac{\pi}{2}.
\end{align}

For a given ansatz, Eq.\ \eqref{eq:omega_asy}, the numerical procedure to find solutions is straightforward. We solve the differential equations \eqref{eq:des_result} using a fourth-order  Runge-Kutta algorithm. For this solution the conditions \eqref{eq:eta1}, \eqref{eq:eta2_detail}, \eqref{eq:psi_taup}, and \eqref{eq:theta_taup} are evaluated. We search for roots using the Powell hybrid method in the GNU scientific library \cite{GSL}. 

Of two different pulses the one with the lower amplitude $V_0$ is preferable in experiment because less
power is needed to realize it. For an experimentally realizable maximum amplitude this implies
that the theoretical pulse with lower amplitude can be made shorter, which is definitely advantageous.
Hence we search for pulses with lower amplitude among the second-order pulses. 
This is done by using an additional coefficient $b_m$ in the ansatz \eqref{eq:omega_asy} and minimizing the amplitude $V_0$  of the resulting solutions by varying this additional coefficient. 

\subsection{First-order pulses}

\begin{figure}
 \includegraphics[width=\columnwidth]{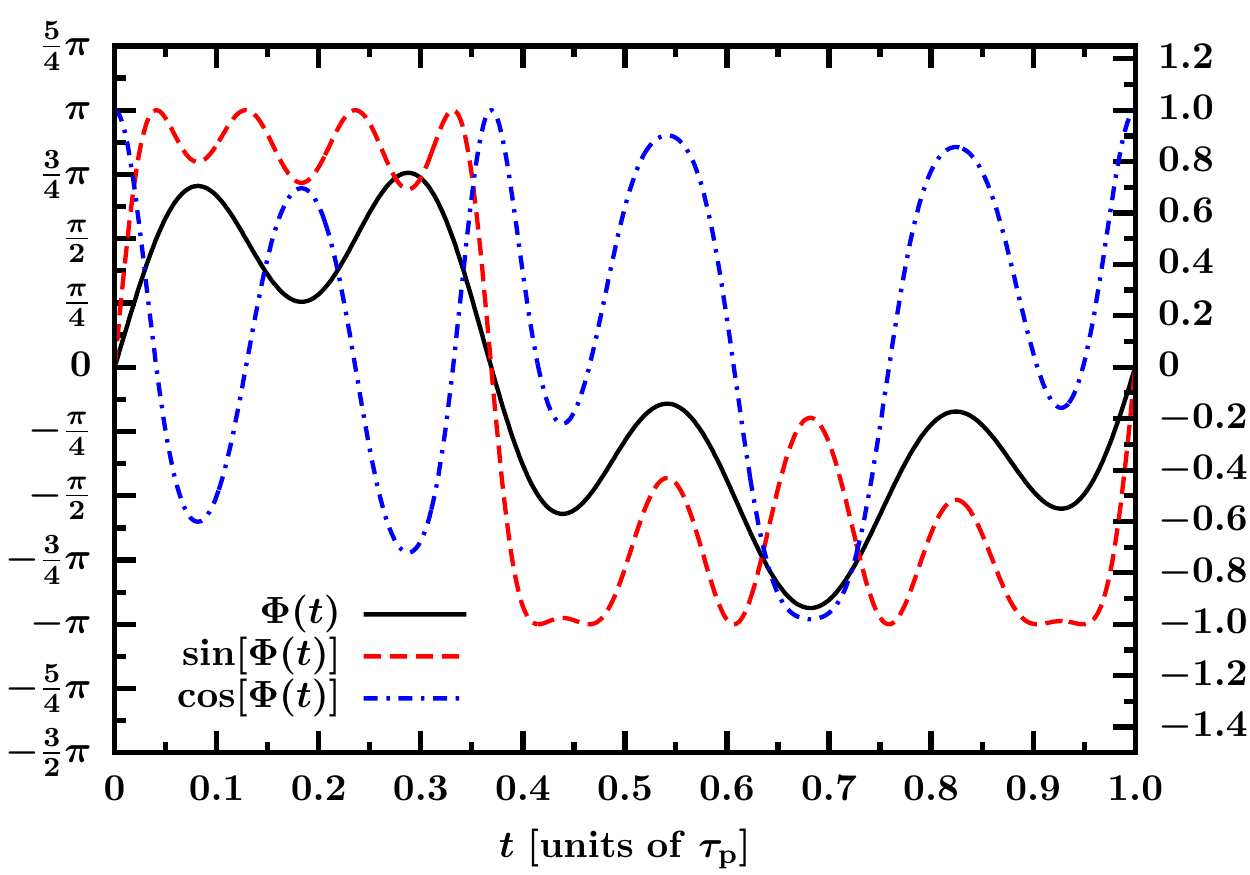}
  \caption{(Color online) Second-order $\pi/2$ pulse FM-2-PI2. We also plot $\sin \Phi(t) \propto v_y(t)$ and $\cos \Phi(t) \propto v_x(t)$ to illustrate the pulse shape in spin space. The left scale refers to $\Phi(t)$ and the right scale to $\sin \Phi(t)$ and $\cos \Phi(t)$, respectively. The Fourier coefficients for this pulse are given in Table \ref{tab.second_order}.
  \label{fig.FC-2ASY-PI2}}
\end{figure}

For first-order pulses and the pure dephasing model, the set of conditions \eqref{eq:eta1} comprises only three equations given in Appendix \ref{app.a}. Adding conditions \eqref{eq:angle} and
\eqref{eq:theta_taup} five parameters are necessary to construct first-order pulses.
One parameter is the amplitude $V_0$ and the others are the coefficients $b_n$
in ansatz \eqref{eq:omega_asy}. The characteristics of the pulses are reported in Table \ref{tab.first_order}.
The pulses are plotted in Figs.\ \ref{fig.FC-1ASY-PI} and \ref{fig.FC-1ASY-PI2}.
Note that the composite and continuous amplitude-modulated pulses found in  Ref.\ \onlinecite{pasin09a} have comparable amplitudes for first-order pulses.

\subsection{Second-order pulses}

\begin{figure}
 \includegraphics[width=\columnwidth]{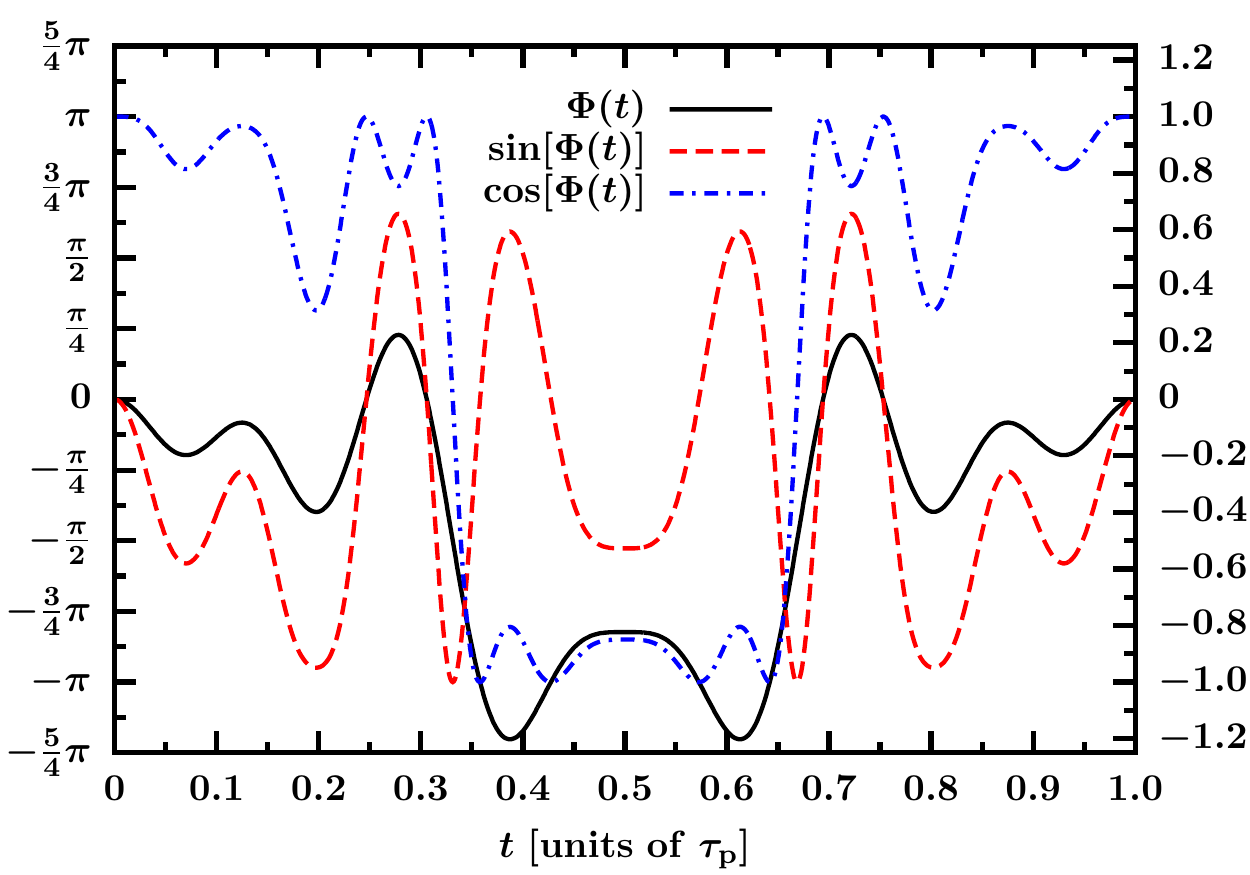}
  \caption{(Color online) Minimized second-order $\pi$ pulse FM-2-MIN-PI; $\sin \Phi(t) \propto v_y(t)$ and $\cos \Phi(t) \propto v_x(t)$ show the pulse shape in spin space. The left scale refers to $\Phi(t)$ and the right scale to $\sin \Phi(t)$ and $\cos \Phi(t)$, respectively. The Fourier coefficients for this pulse are given in Table \ref{tab.second_order_min}.
  \label{fig.FC-2ASY-MIN-PI}}
\end{figure}

Second-order pulses additionally have to fulfill conditions \eqref{eq:eta2_detail}. These equations again simplify for a purely dephasing bath leading to six additional integral conditions besides the first-order terms
(see Appendix \ref{app.b}). Note that more equations are to be fulfilled than 
for amplitude modulation \cite{pasin09a} because the frequency-modulated pulses involve all three spin
directions.
The solutions for $\pi$ and $\pi/2$ pulses are given in Table \ref{tab.second_order} and they are displayed in Figs.\ \ref{fig.FC-2ASY-PI} and \ref{fig.FC-2ASY-PI2}.
Numerically, the double integrals in Eqs.\ \eqref{eq:double_integral} are particularly demanding.
Full quantum mechanical studies of higher-order pulses will be hampered by even higher
dimensional integrals occurring in the Magnus expansion \cite{MagnusExp}.
An alternative route, which may be numerically more efficient, consists of the direct
solution of the Schr\"odinger equation \cite{sengu05}. The mathematical 
existence of higher-order pulses is known \cite{khodj10}.

\begin{figure}
 \includegraphics[width=\columnwidth]{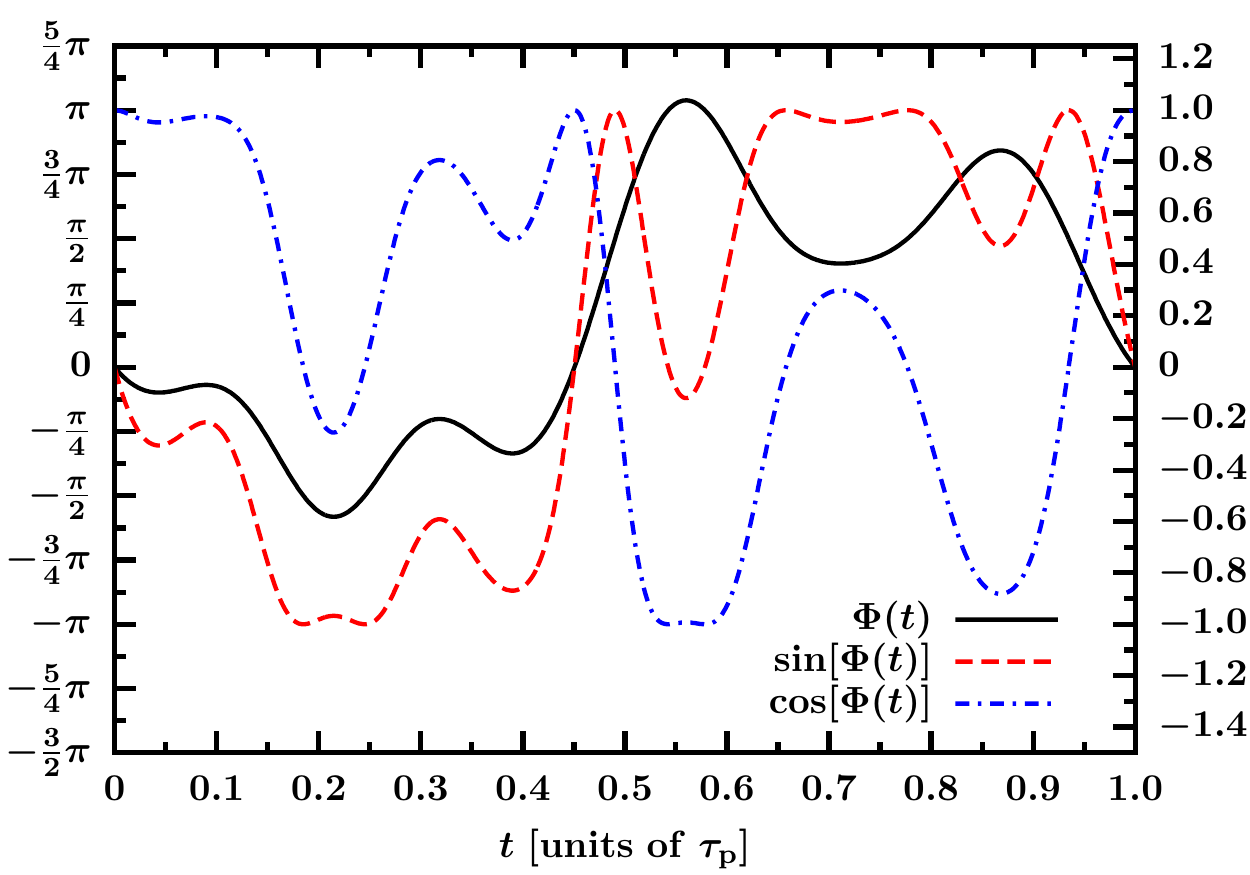}
  \caption{(Color online) Minimized second-order $\pi/2$ pulse FM-2-MIN-PI2. We also plot $\sin \Phi(t) \propto v_y(t)$ and $\cos \Phi(t) \propto v_x(t)$ to illustrate the pulse shape in spin space. The left scale refers to $\Phi(t)$ and the right scale to $\sin \Phi(t)$ and $\cos \Phi(t)$, respectively. The Fourier coefficients for this pulse are given in Table \ref{tab.second_order_min}.
  \label{fig.FC-2ASY-MIN-PI2}}
\end{figure}

Since we are interested in pulses with low amplitudes, we aim at minimizing the amplitude.
To this end, we add another  Fourier coefficient to the ansatz \eqref{eq:omega_asy} and
vary this additional parameter. In this way, we obtained the pulses FM-2-MIN-PI and FM-2-MIN-PI2 
given in Table \ref{tab.second_order_min} and plotted in Figs.\ \ref{fig.FC-2ASY-MIN-PI} and  \ref{fig.FC-2ASY-MIN-PI2}. 
Empirically it turned out to be more efficient to consider $b_{14}$ instead of $b_{11}$ as an additional coefficient. It is expected that even lower amplitudes can be achieved by using further coefficients. But our calculations with different coefficients, not shown here, indicate that 
this route would improve the amplitude only by $1-2\%$ at the expense of a more complex pulse shape. 

By using only one free coefficient ($b_{14}$) we found $\pi/2$ pulses with amplitudes lower than $9.0/\tau_\mathrm{p}$ to be compared with the amplitude-modulated pulses \cite{pasin09a} with amplitude $11.5/\tau_\text{p}$. For $\pi$ pulses we need $V_0=10.7/\tau_\mathrm{p}$ in comparison to 
$11.0/\tau_\text{p}$ for amplitude modulation. The amplitudes of the amplitude-modulated pulses refer
to piecewise constant  pulses; for continuous pulses they are even higher.
Of course, the reduction of the amplitudes for frequency-modulated pulses is not spectacular. But
it is remarkable that pulses with relatively low amplitudes can be found despite the
larger number of conditions to be fulfilled: the frequency-modulated pulses fulfills $9+2=11$ equations
including the conditions for the angles, and the amplitude-modulated pulse fulfills $5+1=6$ conditions
including the condition for the rotation angle.

\section{Conclusions} 
\label{sec:conclusion}

In this paper we extended the existing perturbative approach to decouple a spin from a quantum mechanical bath by means of short control pulses in two ways. 

First, we allowed for a time-dependent bath, which means both
the bath Hamilton operator and the coupling operators may have an explicit, analytical
time dependence. Yet, we found that this time dependence does not alter the
requirements for the pulse shape which were derived previously for time-independent baths \cite{pasin09a}. 
Hence, the pulses found previously are also applicable for time-dependent environments
as they arise, for instance, in time-dependent reference frames or in the interaction picture of otherwise
time-independent Hamiltonians. This is our first key result.

Second, we studied frequency-modulated pulses in first order and in second order in the
pulse duration $\tau_\text{p}$ for quantum mechanical baths. Previously, only
amplitude modulation was considered explicitly for quantum mechanical baths \cite{pryad08a,pryad08b,pasin09a}.
Frequency modulation was so far studied for static baths only \cite{skinn06}.
We provide explicit solutions for continuous frequency-modulated pulses with amplitudes which
have been minimized empirically.
Such pulses are expected to be useful in experiments where no amplitude modulation can be realized or
where the achievable accuracy for frequency modulation is superior to the accuracy of amplitude modulation.
For instance, they can be used to implement realistic optimized dynamic decoupling \cite{uhrig10a,pasin11a}
where the dynamic decoupling sequence is adapted to pulses of finite length.
The frequency-modulated pulses constitute our second key result.

We emphasize that modulated pulses correspond to quantum gates which are robust against decoherence in the framework of quantum-information processing, e.g., the $\pi/2$ pulse about $\sigma_y$ preceded by a $\pi$
pulse about $\sigma_z$ realizes  the important Hadamard gate up to a global factor $i$.
\footnote{This fact was stated in Refs.\ \onlinecite{pasin08a} and  \onlinecite{pasin08b} in a too shortened
way leaving out the $\pi$ pulse.}.

Further work should concentrate on higher-order terms not studied here. Such terms comprise higher-dimensional integrals so that the numerical effort increases considerably. 
Another promising route is to extend  the model from pure dephasing to general decoherence. 
This would allow for systems with finite $T_1$ as well, at the expense of more complex
pulses. 

But at the present stage, it is also called for to verify the performance
of the proposed pulses experimentally in order to assess how promising further
extensions would be.

\begin{acknowledgments}
We thank Christopher Stihl, Nils Drescher, Frederik Keim, and Leonid Pryadko for useful discussions and comments. The study of frequency modulation was triggered by a discussion with Michael Biercuk
and Hermann Uys. We acknowledge financial support of the DFG under Project UH 90/5-1.
\end{acknowledgments}

\appendix

\section{First-order conditions}
\label{app.a}

For the first-order conditions we inspect Eq.\ \eqref{eq:eta1} to find the corresponding scalar equations.
In a purely dephasing bath the sum over $j$ collapses to $j=z$, resulting in the scalar equations
\begin{subequations}
\begin{align}
\eta_{11} &= \int_0^{\tau_\mathrm{p}} a_y(t) \sin(\psi(t)) - (1- \cos(\psi)) a_x(t) a_z(t) \mathrm{d} t    \\
\eta_{12} &= \int_0^{\tau_\mathrm{p}} a_x(t) \sin(\psi(t)) + (1- \cos(\psi)) a_y(t) a_z(t) \mathrm{d} t    \\
\eta_{13} &= \int_0^{\tau_\mathrm{p}} \cos(\psi(t)) + (1- \cos(\psi))  {a_z(t)}^2 \mathrm{d} t     . 
\end{align}
\end{subequations}
The $a_i$ are the components of the global axis of rotation parametrized in Eq.\ \eqref{eq:a_para}.

\section{Second-order conditions}
\label{app.b}

For the second-order conditions, we additionally have to consider Eqs.\ \eqref{eq:eta2_detail} to find the corresponding scalar equations.
Again certain sums collapse due to the purely dephasing bath model and we eventually obtain
\begin{subequations}
\begin{align}
\eta_{21} &= \int_0^{\tau_\mathrm{p}} t \left[ a_y(t) \sin(\psi(t)) - (1- \cos(\psi)) a_x(t) a_z(t) \right] \mathrm{d} t  \\
\eta_{22} &= \int_0^{\tau_\mathrm{p}} t \left[ a_x(t)\sin(\psi(t)) + (1- \cos(\psi)) a_y(t) a_z(t)  \right] \mathrm{d} t   \\\
\eta_{23} &= \int_0^{\tau_\mathrm{p}} t \left[ \cos(\psi(t)) + (1- \cos(\psi))  {a_z(t)}^2  \right] \mathrm{d} t  
\end{align}
\end{subequations}
and
\begin{subequations}
\label{eq:double_integral}
\begin{align}
\eta_{24} &= \int_0^{\tau_\mathrm{p}}  \mathrm{d} t_1 \int_0^{t_1} \mathrm{d} t_2  \left[ n_{yz}(t_1)n_{zz}(t_2) - n_{zz}(t_1)n_{yz}(t_2) \right]   \\
\eta_{25} &= \int_0^{\tau_\mathrm{p}}  \mathrm{d} t_1 \int_0^{t_1} \mathrm{d} t_2  \left[ n_{zz}(t_1)n_{xz}(t_2) - n_{xz}(t_1)n_{zz}(t_2) \right]  \\
\eta_{26} &= \int_0^{\tau_\mathrm{p}}  \mathrm{d} t_1 \int_0^{t_1} \mathrm{d} t_2  \left[ n_{xz}(t_1)n_{yz}(t_2) - n_{yz}(t_1)n_{xz}(t_2) \right] . 
\end{align}
\end{subequations}
The matrix elements $n_{ij}(t)$ occurring here are those of the rotation matrix
$D_{\hat{a}}(-\psi)$ given explicitly in Eq.\ \eqref{eq:matrix}. The components $a_i$ are parametrized in Eq.\ \eqref{eq:a_para}.

\section{Rotation matrix}
\label{app.c}

To derive the matrix $D_{\hat{a}}(-\psi)$ we refer the reader to Ref.\ \onlinecite{pasin09a}. It is calculated by comparison of the coefficients in Eq.\ \eqref{eq:matrixcoeff}. We obtain the matrix \eqref{eq:matrix} below, where the time dependencies of $\psi(t)$ and $\hat{a}(t)$ are omitted for clarity:
\begin{widetext}
\begin{align}
D_{\hat{a}}(-\psi) = \left(\begin{array}{ccc}
\cos \psi + (1- \cos \psi)a^2_{x} 	   & a_{z} \sin \psi + (1 - \cos \psi) a_{x} a_{y}  & -a_{y} \sin \psi + (1 - \cos \psi) a_{x} a_{z} \\
-a_{z} \sin \psi + (1- \cos \psi)a_{x} a_{y} &     \cos \psi + (1 - \cos \psi) a^2_{y}    & a_{x} \sin \psi + (1 - \cos \psi) a_{y} a_{z} \\
a_{y} \sin \psi + (1- \cos \psi)a_{x} a_{z}  & -a_{x} \sin \psi + (1 - \cos \psi) a_{y} a_{z} & \cos \psi + (1 - \cos \psi) a^2_{z} 
                      \end{array}\right) 
\label{eq:matrix}
\end{align}
\end{widetext}

%\bibliographystyle{apsrev4-1}
%\bibliography{pulses3}

%

\end{document}